\documentclass[a4paper,11pt]{article}
\usepackage{pos}
\usepackage{amsmath}
\usepackage{amsfonts}
\usepackage{amsthm}
\usepackage{mathtools}
\usepackage{dsfont}
\usepackage{bm}
\usepackage{bbm}
\usepackage{graphicx}
\usepackage[capitalise]{cleveref}
\usepackage{hyperref}
\usepackage[normalem]{ulem}
\usepackage{yfonts}

 \newcommand{\lsim}{{\;\raise0.3ex\hbox{$<$\kern-0.75em\raise-1.1ex\hbox{$\sim$}}\;}}
\newcommand{\gsim}{{\;\raise0.3ex\hbox{$>$\kern-0.75em\raise-1.1ex\hbox{$\sim$}}\;}}
\def\bea{\begin{eqnarray}}
\def\eea{\end{eqnarray}}
\def\bec{\begin{center}}
\def\ec{\end{center}}

\def\beq{\begin{equation}}
\def\eeq{\end{equation}}

\def\bea{\begin{eqnarray}}
\def\eea{\end{eqnarray}}
\def\beq#1\eeq{\begin{align}#1\end{align}}
\def\beqnn#1\eeq{\begin{align*}#1\end{align*}}
\def\ba{\begin{array}}
\def\ea{\end{array}}
\def\bc{\begin{center}}
\def\ec{\end{center}}

\def\l{\lambda}

\newcommand{\alphap}{\alpha^\prime}
\renewcommand{\epsilon}{\varepsilon}

\newcommand{\mc}{\mathcal}

\newcommand{\coma}{\, , \quad}
\newcommand{\fstop}{\, .}
\def\mpl{M_{P}}
\def\im{{\text{Im} \,}}

\title{Axion Theory and Model Building}

\author[a]{Kiwoon Choi}
\author[b]{Nicole Righi}

\affiliation[a]{Particle Theory and Cosmology Group, Center for Theoretical Physics of the Universe,\\
  Institute for Basic Science (IBS), Daejeon 34126, South Korea}

\affiliation[b]{Theoretical Particle Physics and Cosmology Group, Physics Department, King's College London,\\ Strand, London WC2R 2LS, United Kingdom}

\emailAdd{kchoi@ibs.re.kr}
\emailAdd{nicole.righi@kcl.ac.uk}

\abstract{Axions are light pseudoscalar bosons postulated with many motivations in particle physics and cosmology, including the strong CP problem and the dark matter in our Universe.  In this lecture notes, we discuss
a variety of known ultraviolet (UV) theories for axions
 and their low energy properties. We are primarily concerned with the
quantum chromodynamics axion solving the strong CP problem, as well as  lighter axion-like particles.
 In regard to their UV origin, such light axions may  arise from the spontaneous breakdown of
a linearly realized global Peccei-Quinn $U(1)$ symmetry in the context  of 4-dimensional effective field theories, or they may originate from  a  gauge field in higher dimensional theories.    
 It is noted that different UV models for these axions predict a distinctive pattern of low energy axion couplings, which may have interesting implications for laboratory, astrophysical, or cosmological studies of  axions. We also provide an introductory discussion of the effective field theory for axions from
$p$-form gauge fields in string theory with concrete examples.}

\FullConference{1st General Meeting and 1st Training School of the COST Action COSMIC WSIPers (COSMICWISPers)\\ 
5-14 September, 2023\\
Bari and Lecce, Italy\\}

\begin{document}
\maketitle

\section{Introduction}

Axions are light pseudo-scalar bosons described by an angular field variable. 
There are many motivations to postulate axions, which have been 
discussed in many excellent reviews and lectures, for instance in \cite{Kim:2008hd,Marsh:2015xka,DiLuzio:2020wdo,Choi:2020rgn,Hook:2018dlk,Reece:2023czb}.
They include first of all  the strong 
CP problem of the Standard Model (SM) of particle physics, which can be solved by
 a specific type of axion called the
quantum chromodynamics (QCD) axion.
Light axions  may also constitute the dark matter or dark radiation in our Universe, even the dark energy for ultralight axion
with a mass near the Hubble scale. Such light axions can have  a variety of  observable  consequences including astrophysical and cosmological phenomena.
Moreover, an axion with certain features could have played a key role in the early Universe inflation.  Finally,
axions generically appear in 4-dimensional effective theories of string/M theory, which is regarded as the best candidate for a theory incorporating both the SM and quantum gravity.  
In this lecture, we discuss some theory and model building aspects of axion physics while focusing on the connection between the ultraviolet (UV) origin of axions and
the associated low energy axion physics.  In the latter part of the lecture, we provide an introductory discussion of the effective  theory for string theory axions with concrete examples.


\section
{QCD axion and axion-like particles} \label{sec:axion_eft}

As is well known,
the SM  involves two CP-odd angle parameters, the Kobayashi-Maskawa phase and the QCD angle\footnote{As for the   angles $\theta_{W}$ and $\theta_Y$ for the electroweak gauge group $SU(2)_W\times U(1)_Y$, one combination  can be rotated away by the baryon or lepton number
($U(1)_{B/L}$) transformation.
The combination $\theta_{\rm EM}= \theta_W+\theta_Y$ is invariant under
$U(1)_{B/L}$, however it is relevant only when the nonzero magnetic flux or topologically non-trivial spacetime are involved \cite{Reece:2023czb}.}, which are given by
\bea \delta_{\rm KM}= \mbox{arg}\cdot\mbox{det}([y_uy_u^\dagger, y_dy_d^\dagger]), \quad \bar\theta =\theta_{\rm QCD}+\mbox{arg}\cdot\mbox{det}(y_uy_d)\label{repara},\eea where
$y_{u,d}$ are the Yukawa coupling matrices of the 3-generations of the up and down-type quarks.
Observed CP violation in 
 the weak interactions implies
$
\delta_{\rm KM}\,\sim \, 1, 
$
while the absence of  CP violation in the strong interactions leads to the upper bound 
$
|\bar \theta|\,\lesssim\, 10^{-10}
$
\cite{Kim:2008hd}.
Such a small value of $\bar\theta$  requires an unnatural fine-tuning of the involved parameters in (\ref{repara}), called the strong CP problem \cite{Kim:2008hd,Marsh:2015xka,DiLuzio:2020wdo,Choi:2020rgn}\footnote{It has been argued that the smallness of $\bar\theta$ can not be understood by an anthropic selection in the multiverse  \cite{Ubaldi:2008nf,Dine:2018glh}. 
See 
\cite{Kaloper:2017fsa} for an anthropic argument which may account for small $\bar\theta$, which was critically re-examined in \cite{Dine:2018glh}.}.

An  appealing solution of the strong CP problem is to introduce 
a global  Peccei-Quinn (PQ) $U(1)$ symmetry \cite{Peccei:1977hh} which is
(i) non-linearly realized in the low energy limit, with the associated  Nambu-Goldstone boson ``the QCD axion'' \cite{Weinberg:1977ma,Wilczek:1977pj},
and
(ii) explicitly broken dominantly by the QCD anomaly.  
For a generic low energy effective theory with non-linear $U(1)_{\rm PQ}$,
one can choose a field basis 
for which \emph{only} the axion field
 transforms under $U(1)_{\rm PQ}$ \cite{Georgi:1986df} as
\bea
U(1)_{\rm PQ}:  \quad \frac{a(x)}{f_a} \,\, \rightarrow \,\, \frac{a(x)}{f_a} +\alpha \quad (\alpha=\mbox{constant})\coma\label{nonlinear_pq}\eea
where $f_a$ is
 the axion decay constant defining the axion field range as
\bea a(x)\cong a(x) +2\pi f_a \label{axion_peri}\fstop\eea
In this field basis which will be called the Georgi-Kaplan-Randall (GKR) basis \cite{Georgi:1986df}, the  generic  axion
effective Lagrangian at scales below $f_a$ (but above the weak scale) 
takes the form
\bea
{\cal L}_{\rm  axion}&=&\frac{1}{2}\partial_\mu a\partial^\mu a+\frac{\partial_\mu a(x)}{f_a}\left[\sum_\psi c_\psi\bar \psi  \bar\sigma^\mu \psi +\sum_\phi  ic_\phi\left(\phi^\dagger D_\mu \phi - {\rm h.c.}\right) \right]\nonumber \\
&+&\frac{1}{32\pi^2}\frac{a(x)}{f_a}\left(\sum_{F^A=G,W,B,\dots} c_AF^{A\mu\nu}\tilde F^A_{\mu\nu}\right)
+\delta {\cal L}_{\rm axion},\coma\label{axion_lag}
\eea
where
 $c_\psi$ and  $c_\phi$ parameterize the
PQ-preserving axion derivative  couplings
to the canonically normalized chiral fermion $\psi$ and scalar field $\phi$, while $c_A=(c_G,c_W,c_B,...)$ parameterize  the
PQ-breaking couplings\footnote{Note that $c_A$ can be regarded as PQ-preserving couplings 
in perturbation theory as $F^A_{\mu\nu}\tilde F^{A\mu\nu}$ is a total divergence.} to the gauge fields $F^A_{\mu\nu}$ including the SM gauge fields $G^a_{\mu\nu}, W^i_{\mu\nu}, B_{\mu\nu}$.
Here  $\delta {\cal L}_{\rm axion}$ can include additional PQ-breaking terms, e.g. an axion potential
induced by Planck scale physics such as quantum gravity \cite{Barr:1992qq,Kamionkowski:1992mf,Holman:1992us}.
For the Lagrangian (\ref{axion_lag}) to be valid over the full axion field range, 
the axion periodicity (\ref{axion_peri}) requires  that $c_A$  are quantized \cite{Reece:2023czb}, e.g. integers for gauge fields
with the kinetic terms  $-\frac{1}{4g_A^2}F^{A\mu\nu}F^A_{\mu\nu}$ and properly normalized gauge couplings $g_A^2$.

To obtain the axion couplings at lower energy scales, 
one may first consider  the renormalization group (RG) evolution of the parameters $c_\psi, c_\phi$ and $c_A$. The quantized nature of $c_A$
implies that $c_A$ are RG-invariant. On the other hand, generically $c_{\phi}$ and $c_\psi$  experience non-trivial RG evolution due to the associated
gauge and Yukawa interactions. For instance, for $\phi$ and $\psi_{1,2}$ having
 the Yukawa coupling
$y\phi\psi_1\psi_2$, one finds \cite{Choi:2017gpf,Chala:2020wvs,Bauer:2020jbp,Choi:2021kuy}
\bea
&& \hskip -0.5cm 
\frac{d c_\phi}{d\ln \mu}= \frac{|y|^2}{8\pi^2}(c_\phi+c_{\psi_1}+c_{\psi_2}), \nonumber \\
&& \hskip -0.5cm 
\frac{d c_{\psi_{1,2}}}{d\ln \mu}= \frac{|y|^2}{16\pi^2}(c_\phi+c_{\psi_1}+c_{\psi_2}) 
-\frac{3}{2} \Big(\frac{g_A^2}{8\pi^2}\Big)^2\mathbb{C}_{A}(\psi_{1,2}) \Big(c_A-2\sum_{\psi=\psi_i} c_{\psi} {\rm tr}(T^2_A(\psi))\Big), 
\eea
where 
 $T_A(\psi)$ is the gauge charge of $\psi$ for the $A$-th gauge group and  $\mathbb{C}_{A}(\psi)$ is its quadratic Casimir.
There can also be threshold corrections to axion couplings when heavy particles are integrated out
\cite{Bauer:2020jbp,Choi:2021kuy}.
Including the relevant RG evolution and threshold corrections,
the axion couplings at 
 $\mu\sim 1$ GeV are given by 
 \bea
 \frac{1}{32\pi^2}\frac{a(x)}{f_a}\Big(c_\gamma F^{\mu\nu}\tilde F_{\mu\nu} + c_G  G^{a\mu\nu}\tilde G^a_{\mu\nu}\Big)
 +\sum_{\Psi=u,d,s,e,\mu} \frac{C_\Psi}{2} \frac{\partial_\mu a}{f_a}\, \bar\Psi\gamma^\mu\gamma_5\Psi-\delta {V}_{\rm axion},
 \eea
 where $F_{\mu\nu}$ is the $U(1)_{\rm em}$ gauge field strength, $\Psi=u,d,s,e,\mu$ are the relevant light Dirac quarks and leptons,
  $\delta V_{\rm axion}$ is the axion potential induced by UV physics such as quantum gravity, and
\bea
c_\gamma &=& c_W+c_B,\nonumber \\
C_u(\mu)&=& c_{q}(v)+c_{u^c}(v)+c_H(v) +\Delta C_u,\nonumber \\
C_d(\mu)&=& c_{q}(v)+c_{d^c}(v)-c_H(v) +\Delta C_d, \,\, \mbox{etc.} 
\eea
Here
 $q,u^c,d^c,$ and $H$ denote the 1st generation of the left-handed $SU(2)_W$-doublet quarks, up and down-type singlet antiquarks, and the Higgs doublet, respectively, $v=246$ GeV is the weak scale, and
  $\Delta C_\Psi$ are the radiative corrections over the scales from  
  $v=246$ GeV to $\mu\sim 1$ GeV.

A nonzero integer  $c_G$ describing the axion coupling to the gluons in (\ref{axion_lag})
represents the PQ-breaking by QCD anomaly. Around the QCD scale, it
generates an axion potential \cite{DiLuzio:2020wdo,villadoro}
\bea
V_{\rm QCD}(a)
\simeq -
\frac{f_\pi^2 m_\pi^2}{m_u+m_d}\sqrt{{m_u^2+m_d^2+2m_um_d\cos(c_Ga/f_a)}},
\eea
where $m_{u,d}$ are the light quark masses and  $f_\pi= 92.4$ MeV is the pion decay constant.
Including this QCD anomaly-induced potential,
the  full axion potential is given by
\bea
V_{\rm axion}= V_{\rm QCD}+\delta V_{\rm axion}.\eea
If $U(1)_{\rm PQ}$ is broken \emph{dominantly} by the QCD anomaly to the extent that 
$\delta V_{\rm axion}/V_{\rm QCD}< 10^{-10}$, 
the axion vacuum value is small enough to solve the strong CP problem as
\bea |\bar\theta| \equiv \frac{|c_G{\langle a(x)\rangle}|}{f_a} 
\,< \,10^{-10}\quad (c_G\neq 0).\eea
 Such an axion is called the QCD axion. Obviously then the axion mass 
 is determined by  the axion coupling to the gluons as
\bea
m_{a_{\rm QCD}} =f_\pi m_\pi\frac{ \sqrt{m_um_d}}{m_u+m_d}\frac{c_G }{f_a}\simeq 0.57\times \Big(\frac{10^{10}\,{\rm GeV}}{f_a/c_G}\Big) \, {\rm meV}. \label{mass_gluon_coupling}\eea
Note that  $c_G$ also parameterizes   the number of degenerate vacua of $V_{\rm QCD}$.  Therefore, for a QCD axion with $|c_G| >1$, the associated cosmic domain walls can cause a severe problem  in the cosmological scenario with post-inflationary PQ phase transition \cite{Marsh:2015xka}. 

Generically there can also be different type of axions dubbed axion-like particles (ALPs) \cite{Marsh:2015xka,Choi:2020rgn}.  They include for instance  a heavy ALP with $\delta V_{\rm axion}(a) \gg f_\pi^2m_\pi^2$, as well as an ultralight ALP with $\delta V_{\rm axion}(a) \ll f_\pi^2m_\pi^2$ and
 $c_G=0$, for which  
\bea
m_{\rm heavy-ALP} \gg    \frac{10^{10}\,{\rm GeV}}{f_a} \, {\rm meV}, \quad
m_{\rm ultralight-ALP} \ll     \frac{10^{10}\,{\rm GeV}}{f_a} \, {\rm meV} . 
\eea

In many cases,
the axion couplings  
relevant for cosmological, astrophysical, or laboratory processes are those  {below the QCD scale}. Those couplings include \cite{chang_choi,villadoro}
\bea
{\cal L}_{\rm eff}&=& \frac{1}{4}g_{a\gamma}a(x)\vec E\cdot \vec B +\sum_{\ell=e,\mu} \frac{1}{2}g_{a\ell}\bar \ell\gamma^\mu\gamma_5 \ell 
\nonumber \\
&+&\frac{1}{2}\partial_\mu a(x)
\Big[ g_{ap} \bar p \gamma^\mu\gamma_5 p 
 + g_{an} \bar n \gamma^\mu \gamma_5 n 
  + \frac{g_{a\pi N}}{f_\pi} \big( i \pi^+ \bar p \gamma^\mu n - i \pi^- \bar n\gamma^\mu p\big)  \Big]
\nonumber \\
&+&\frac{1}{2}\frac{g_{a\pi}}{f_\pi}\partial^\mu a(x)\Big( \pi^0 \pi^+ \partial_\mu\pi^- +\pi^0\pi^-\partial_\mu\pi^+ - 2\pi^+\pi^-
\partial_\mu\pi^0\Big),
\label{low_eff}
\eea
where 
\bea \label{Coeffs}
&& 
\hskip -1cm
g_{a \gamma} =  \frac{\alpha_{\rm em}}{2\pi f_a}\Big(c_W+c_B -\frac{2}{3} \frac{(4m_d+m_u)}{(m_u+m_d)}c_G \Big),
\quad
g_{ap}- g_{an} = \frac{1.27}{f_a} \left(C_u - C_d + \Big(\frac{m_u - m_d}{m_u+ m_d} \Big)c_G\right), \nonumber \\
 &&\hskip -1cm 
g_{ap} + g_{an}= \frac{0.52}{f_a} \Big(C_u + C_d- c_G \Big), \quad 
 g_{a\pi} = \frac{2\sqrt{2}}{3}g_{a\pi N}  =\frac{g_{ap} -g_{an}}{1.9}, \quad g_{a\ell}=\frac{C_\ell}{f_a}
\eea 
for $C_{u,d}$ renormalized at $\mu=2$ GeV. Then,
putting the known light quark mass ratios, one finds
\bea
&&\hskip -1cm g_{a\gamma}\simeq \frac{\alpha_{\rm em}}{2\pi}\frac{1}{f_a} \Big(c_W+c_B -1.92 c_G\Big),
\label{axion_photon} \\
&& \hskip -1cm g_{ap}\simeq \frac{1}{f_a}\Big(0.90 C_u -0.38 C_d-0.48 c_G \Big), \quad g_{an}\simeq \frac{1}{f_a}\Big( 0.90 C_d-0.38 C_u -0.04 c_G\Big). \label{axion_nucleon}
 \eea

 An important quantity for experimental detection of axions is the coupling to mass ratio $g_{a\gamma}/m_a$.  For a QCD axion, (\ref{mass_gluon_coupling}) and (\ref{axion_photon}) imply
 \bea
\mbox{QCD axion:}\quad \frac{g_{a\gamma}}{m_{a}} \sim \frac{\alpha_{\rm em}}{2\pi}\frac{1}{f_\pi m_\pi}\frac{c_W+c_B}{c_G}.\eea
 Typical QCD axion models give $c_{W,B}\sim c_G$, therefore predict   $g_{a\gamma}/m_a\sim{\alpha_{\rm em}}/2\pi f_\pi m_\pi$. This
  corresponds to  \emph{the QCD axion band}  on the plane spanned by 
  $(m_a,g_{a\gamma})$,  which is the primary target of axion search experiments \cite{Kim:2008hd}.   
 Yet there can be models giving $c_{W,B}\gg c_G$ \cite{Agrawal:2017cmd},  most notably
 the clockwork axion model \cite{cw_axion1,cw_axion2,cw_axion3} with multiple ($N>1$) axions, in which $c_{W,B}/c_G\sim q^{N-1}$ for an integer-valued model parameter $q$. 
 For  ALPs with nonzero coupling to the photon, one typically finds
\bea
\mbox{Heavy ALP:} \quad \frac{g_{a\gamma}}{m_{a}} \,\ll \, \frac{\alpha_{\rm em}}{2\pi}\frac{1}{f_\pi m_\pi}, \qquad
\mbox{Ultralight ALP:} \quad \frac{g_{a\gamma}}{m_{a}}\, \gg\,  \frac{\alpha_{\rm em}}{2\pi}\frac{1}{f_\pi m_\pi},\eea
Note that clockwork QCD axion and ultralight ALP can have a similar value of
$g_{a\gamma}/m_a$ \cite{cw_axion3}.


 \section{Axion models} \label{sec:model}

As for the UV origin of axions, one can consider  two types of models.
The first  are 4-dimensional (4D) models with  a \emph{linearly-realized} $U(1)_{\rm PQ}$ symmetry which is spontaneously broken at  a scale around $f_a$ \cite{Peccei:1977hh,Weinberg:1977ma,Wilczek:1977pj,ksvz1,ksvz2,dfsz1,dfsz2}. The second  are higher-dimensional models involving a $p$-form gauge field, in which  axions arise as the zero modes of $p$-form gauge field at the compactification scale \cite{string_axion1,string_axion2,5d_axion1,5d_axion2}. For the second type,  there is no notion of linear $U(1)_{\rm PQ}$ symmetry, and the non-linear  $U(1)_{\rm PQ}$ in 4D effective theory can be identified as a \emph{locally well-defined, but globally ill-defined gauge symmetry} in the underlying higher-dimensional theory.  Due to this, PQ-breaking by quantum gravity in the second type models can be exponentially suppressed in an appropriate limit. In the following, we briefly discuss 
typical examples of both types of axion models.
One of our primary concerns will be the pattern of low energy axion couplings predicted by these models. 

\subsection{Models
   with  a linearly realized PQ symmetry}

   Generic linear PQ symmetry  is defined as
 \bea
 U(1)_{\rm PQ}:   \,\, \Phi \rightarrow  e^{iq_\Phi \alpha}\Phi \label{linear_pq}
\eea
with the quantized PQ charges $q_\Phi$ of the matter fields $\Phi=(\phi,\psi)$ in the model.  
The associated Noether current  is given by
\bea
J^\mu_{\rm PQ}=\sum_\Phi\frac{\partial {\cal L}}{\partial (\partial_\mu \Phi)}\frac{\delta \Phi}{\delta\alpha}=-\sum_\psi q_\psi \bar\psi
\bar\sigma^\mu\psi - i\sum_\phi q_\phi (\phi^\dagger D^\mu\phi -{\rm h.c.}),  \label{pq_current}
\eea
and the anomalous variation of the path integral measure of $\psi$ under $U(1)_{\rm PQ}$ leads to  \cite{fujikawa}
\bea
\partial_\mu J^\mu_{\rm PQ}&=&\frac{1}{32\pi^2}\sum_A c_A F^A\tilde F^A\quad \Big(c_A =-2\sum_\psi q_\psi {\rm tr}(T^2_A(\psi)) \Big). \label{pq_divergence}
\eea
In the following, we will take the PQ charge normalization for which
the axion field $a(x)$ parameterizing the vacuum manifold of spontaneously broken $U(1)_{\rm PQ}$ transforms under  $U(1)_{\rm PQ}$ as in 
(\ref{nonlinear_pq}). 
Here we  present three distinct models with linear $U(1)_{\rm PQ}$, which have been widely discussed in the literatures, i.e. Kim-Shifman-Vainshtein-Zakharov (KSVZ) 
 model \cite{ksvz1,ksvz2}, Dine-Fischler-Srednicki-Zhitnitsky  (DFSZ) model \cite{dfsz1,dfsz2}, and composite axion model \cite{composite1,composite2,composite3}.

\subsubsection{KSVZ model}

In KSVZ model \cite{ksvz1,ksvz2},  all SM fields are neutral under $U(1)_{\rm PQ}$. Yet there exist  exotic PQ-charged and gauge-charged fermions which obtain a heavy mass due to the spontaneous breakdown of  $U(1)_{\rm PQ}$.
For illustration, let us consider a simple example with  exotic $SU(3)_c$ (anti)triplet and $SU(2)_W$ singlet left-handed fermion $Q$ ($Q^c$) carrying the $U(1)_Y$ hypercharge $Y_Q$ ($-Y_Q$). Introducing also a PQ-charged, but gauge-singlet, scalar field $\sigma$
for the spontaneous breakdown of $U(1)_{\rm PQ}$,
 the Lagrangian of the PQ sector  is given by 
\bea
{\cal L}_{\rm KSVZ}=\partial_\mu\sigma\partial^\mu\sigma^* 
+ i\bar Q \bar\sigma^\mu D_\mu Q +
i\bar Q^c\bar\sigma^\mu D_\mu Q^c
-\Big(y \sigma QQ^c+{\rm h.c.}\Big) -\lambda \Big(\sigma\sigma^*-\frac{1}{2}f_a^2\Big)^2\label{ksvz}.\eea
The model  is invariant (at tree level) under the  linear PQ symmetry (\ref{linear_pq}) with 
$q_\sigma=1$, $q_Q=q_{Q^c}=-1/2$ in our PQ charge normalization convention, for which
$\partial_\mu J^\mu_{\rm PQ}
=
\frac{1}{32\pi^2}\big(G\tilde G+    3Y_Q^2B\tilde B\big)$.
The vacuum manifold of  spontaneously broken $U(1)_{\rm PQ}$
is described by
\bea \langle \sigma\rangle =\frac{1}{\sqrt{2}}f_ae^{ia(x)/f_a} \label{axion_phase}\eea
with $a(x)\cong a(x)+2\pi f_a$ which transforms under $U(1)_{\rm PQ}$ as (\ref{nonlinear_pq}).

To obtain the low energy axion couplings in  KSVZ model,  one may first replace $\sigma$ with the axion-dependent
vacuum value (\ref{axion_phase}). To go to the GKR basis, one subsequently makes the axion-dependent field redefinition
\bea
Q\rightarrow e^{iq_Qa(x)/f_a}Q, \quad  Q^c\rightarrow e^{iq_{Q^c}a(x)/f_a}Q^c, \label{field_redef}\eea  yielding  the axion effective Lagrangian (\ref{axion_lag}) with
 \bea
 c_{Q, Q^c}=-q_{Q, Q^c}=\frac{1}{2}, \quad
 c_A =(1, 0,  3Y_Q^2) \quad (A=G,W,B).
 \eea
 Note that the axion couplings to gauge fields with the coefficients  $c_A$ 
 arise from the anomalous variation of the path integral measure of $Q,Q^c$ under the 
 field redefinition (\ref{field_redef}) \cite{fujikawa}. As a consequence,  $c_A$ in the GKR basis are identical to the anomaly coefficients
 in  $\partial_\mu J^\mu_{\rm PQ}$.
 

A key feature of the KSVZ axion is that at tree level the axion couplings to the SM fermions and Higgs field are all vanishing.
As a consequence, the  low energy axion couplings to the light quarks and electron are determined mainly by
the RG running  due to the SM gauge interactions \cite{srednicki,chang_choi} and the top quark Yukawa coupling $y_t$ 
\cite{Choi:2017gpf,Bauer:2020jbp}, which occurs at  scales below $m_Q=yf_a/\sqrt{2}$. Specifically one finds  \cite{Choi:2021kuy}
\bea
C_{u,d} (\mu=2 \,{\rm GeV})= r^G_{u,d}c_G+ r^W_{u,d}c_W+r^B_{u,d}c_B,
\quad C_e(m_e)= r^G_{e}c_G+ r^W_{e}c_W+r^B_{e}c_B,
\eea
where the coefficients  $r^A_{u,d,e}$ $(A=G,W,B)$  are estimated   for $m_Q=10^7-10^{16}$ GeV as
\bea
&&  \hskip -0.7cm r^G_{u,d}\simeq 2\times 10^{-2},   \quad r^W_u\simeq (2-5)\times 10^{-4}, \quad r^W_d\simeq (3-8)\times 10^{-4},\quad  r^B_u\simeq (2-6)\times 10^{-5},\nonumber \\
&&
\hskip -0.7cm  r^B_{d}\simeq (1-4)\times 10^{-5}, \quad
  r^G_e\simeq (0.5-1)\times 10^{-3},\quad 
 r^W_e\simeq (4-9)\times 10^{-4}, \quad r^B_e\simeq (1-3)\times 10^{-4}.\nonumber\eea

\subsubsection{DFSZ model}

In DFSZ model \cite{dfsz1,dfsz2}, the  SM fermions are charged under the linear $U(1)_{\rm PQ}$. Again, introducing a PQ-charged gauge-singlet scalar $\sigma$,
the Lagrangian of the minimal DFSZ model is given by
\bea
&&\hskip -1cm {\cal L}_{\rm DFSZ}=\partial_\mu\sigma\partial^\mu\sigma^* 
+ D_\mu H_u^\dagger D^\mu H_u +D_\mu H_d^\dagger D^\mu H_u +\Big(y_u H_uq u^c + y_d H_d q d^c +y_\ell H_d\ell e^c+{\rm h.c.}\Big)
\nonumber \\
&& \hskip 0.3cm 
-\lambda \Big(\sigma\sigma^*-\frac{1}{2}f_a^2\Big)^2
-\Big(\kappa H_uH_d \sigma^2 +{\rm h.c.}\Big)+ \dots,\label{dfsz}\eea
where only the relevant terms are explicitly written. The associated PQ charges  are 
$q_\sigma=1$, $q_{H_u,H_d}=-1$,  $q_{\psi_{\rm SM}}=1/2$, where $\psi_{\rm SM}$ denotes the left-handed SM quarks and leptons, yielding
$\partial_\mu J^\mu_{\rm PQ}= -\frac{1}{32\pi^2}\big(6G\tilde G+ 6W\tilde W +    10B\tilde B\big)$.
Upon ignoring the small corrections of ${\cal O}(|H_{u,d}|^2/f_a^2)$, the vacuum manifold of spontaneously broken $U(1)_{\rm PQ}$ in DFSZ model is described by 
\bea\langle \sigma\rangle =\frac{1}{\sqrt{2}}f_ae^{ia(x)/f_a}\eea for $a(x)\cong a(x)+2\pi f_a$ transforming under $U(1)_{\rm PQ}$ as (\ref{nonlinear_pq}).  

 To obtain the axion couplings in the GKR basis,  one can replace $\sigma$ with its axion-dependent vacuum value and subsequently make the field redefinition
\bea
H_{u,d}\rightarrow e^{iq_{H_{u,d}}a(x)/f_a}H_{u,d},  \quad \psi_{\rm SM}\rightarrow e^{iq_{\psi_{\rm SM}} a(x)/f_a}\psi_{\rm SM}.
\eea
This results in the axion couplings at $\mu\sim f_a$, determined by the coefficients
 \bea
 c_{H_{u,d}} =-q_{H_{u,d}}=1, \quad c_{\psi_{\rm SM}} =-q_{\psi_{\rm SM}}=-\frac{1}{2},  \quad c_A =-(6, 6, 10) \quad (A=G,W,B) ,\eea
 where $c_A$ 
 arise from the anomalous variation of the path integral measure of $\psi_{\rm SM}$ \cite{fujikawa}. 
 
Let $m_{\tilde H}$ denote the mass of the heavier combination of $H_u$ and $H_d$. At $\mu >m_{\tilde H}$, there is no RG running of the axion couplings in DFSZ model.
Once $\tilde H$ is integrated out at $\mu\sim m_{\tilde H}$, $H_{u,d}$ can be parameterized as 
 $H_u = H \sin\beta$  and  $H_d= H^*\cos\beta$, where $H$ is the SM Higgs doublet and $\tan\beta = \langle H_u\rangle /\langle H_d\rangle$. One then finds
 $c_H (\mu = m_{\tilde H})  = c_{H_u}\sin^2\beta-c_{H_d}\cos^2\beta = -\cos 2\beta$. In DFSZ model,
RG running due to $y_t$ begins at
$\mu = m_{\tilde H}$, while the RG running  due to the SM gauge interactions begins at $m_t$ \cite{Choi:2021kuy}. However,  those  radiative corrections can be ignored over the majority of parameter space, giving the low energy couplings at $\mu\sim 1$ GeV as
\bea
C_u \simeq  -2\cos^2\beta, \quad
C_d  \simeq 
C_e \simeq -2\sin^2\beta.\eea 
For the parameter region of the DFSZ model where the RG running effect gives a significant consequence, see \cite{nardi}.

\subsubsection{Composite axion model}

A key motivation for the composite axion is to generate  $f_a$ dynamically without causing a scale hierarchy problem \cite{composite1,composite2}.
The first composite axion model was proposed in \cite{composite1}, involving a confining \emph{axicolor} gauge group $SU(N_a)$ with the 
axicolored left-handed fermions:
\bea
\psi_A=\big[\psi_{A_1},\psi_{A_2}\big]=
\big[(N_a, 3),(N_a,1)\big], \quad \psi^c_{A}=\big[ \psi^c_{A_1}, \psi^c_{A_2}\big]=\big[(\bar N_a, \bar 3), (\bar N_a,1)\big],
\eea
where $N_a,\bar N_a$ denote the $SU(N_a)$ representation,  and $3,\bar 3$ are the $SU(3)_c$ representation. The PQ charges
in our normalization convention are   $q_{\psi_{A_1}}=q_{\psi^c_{A_1}}=1/2$ and $q_{\psi_{A_2}}=q_{\psi^c_{A_2}}=-3/2$, for which 
$\partial_\mu J^\mu_{\rm PQ}= -\frac{N_a}{32\pi^2}G\tilde G.$
Then the confining axicolor gauge interactions 
form 
fermion condensations which break $U(1)_{\rm PQ}$  spontaneously, 
\bea
\langle \psi_{A_1}\psi^c_{B_1}\rangle=  \Lambda_a^3 e^{ia(x)/f_a},\quad \langle \psi_{A_2}\psi^c_{B_2}\rangle =\Lambda_a^3 e^{-3ia(x)/f_a},
\eea
where $f_a\sim \Lambda_a$ and  the periodic axion $a(x)\cong a(x)+2\pi f_a$ transforms under $U(1)_{\rm PQ}$
as (\ref{nonlinear_pq}).

 Recently it has been noted that  in certain composite axion models, $U(1)_{\rm PQ}$ appears
as an accidental symmetry which is valid up to the operators of ${\rm dim}=8$
due to the gauge symmetries of the model  \cite{composite3}.
As a consequence,  $U(1)_{\rm PQ}$ is protected from quantum gravity well enough to implement the axion solution to the strong CP problem. 
The axicolor gauge group is $SU(5)_a$ with the
axicolored left-handed fermions
$\psi_{10}=\big[\psi_{(10, 3)} , \psi_{(10,\bar 3)}\big]$ and  $\psi_{\bar 5}=\big[\psi_{(\bar 5, 3)}, \psi_{(\bar 5,\bar 3)}\big]$, where the subscripts denote the $SU(5)_a\times SU(3)_c$ representation.  
The PQ charges in our normalization convention  are 
$q_{\psi_{10}}=1/10$ and  $q_{\psi_{\bar 5}}=-3/10$, for which $\partial_\mu J^\mu_{\rm PQ}=  \frac{2}{32\pi^2}G\tilde G$.
This $U(1)_{\rm PQ}$ is spontaneously broken by the ${\rm dim}=9$  fermion condensation involving the axion field as
\bea
\langle \psi_{10}\cdot\psi_{\bar 5}\cdot\psi_{\bar 5} \cdot \psi_{10}\cdot \psi_{\bar 5}\cdot \psi_{\bar 5}\rangle=\Lambda_a^9 e^{ia(x)/f_a}.
\eea
As the SM fields have vanishing PQ-charges in these composite axion models, the low energy axion couplings 
 are similar to those of the KSVZ axion.

 \subsection{Axions from higher-dimensional gauge field}
 
In the previous subsection, we presented several models with a linear $U(1)_{\rm PQ}$ which might be simply assumed or may arise as an accidental symmetry of the model. In this subsection, we consider a different type of models in which a 4D axion originates
from  a higher-dimensional gauge field \cite{string_axion1,string_axion2,5d_axion1,5d_axion2}. Such models 
do not admit  a linear
$U(1)_{\rm PQ}$, but  yet have a nonlinear $U(1)_{\rm PQ}$
in the low energy limit, which is intriguingly related to the higher-dimensional gauge symmetry of the model.
We first present a simple 5D model  whose axion shares many features with the axions from $p$-form gauge fields in string theory, 
and later discuss axions in string theory.

\subsubsection
{Axion from  5D gauge field}
 
 Our example is the model proposed in \cite{5d_axion2}. The 5D action of the model is given by
 \bea
 &&\hskip -1cm
 S_{\rm 5D}=\int d^5x \sqrt{-\tilde g}  \,\Big[\,\frac{1}{2}M_5^3 {\cal R}_5(\tilde g) -\frac{1}{4g_{5A}^2}A^{MN}A_{MN} -
 \frac{1}{4g_{5S}^2}G^{aMN}G^a_{MN}\nonumber \\
 &&\hskip -0.5cm +\,i\sum_I\bar Q_I \big( \gamma^MD_M+\mu_I A_{MN}\gamma^{MN}\big) Q_I+ \frac{k_{\rm CS}}{32\pi^2}\frac{\epsilon^{MNPQR}}{\sqrt{-\tilde g}}A_MG^a_{NP}G^a_{QR} + \dots
 \Big], \label{5d_model}
 \eea
 where ${\cal R}_5$ is the Ricci scalar for the 5D metic $\tilde g_{MN}$, $A_{MN}$ and $G^a_{MN}$ are the
 $U(1)_A\times SU(3)_c$  gauge field strength tensors with the 5D gauge couplings $g_{5A}$ and $g_{5S}$, and $Q_I=(Q_i, Q_i^c)$ are $SU(3)_c$-charged,  but $U(1)_A$-neutral  5D (anti)quarks with the  $U(1)_A$  dipole moment $\mu_I$. Here $M_5$ is the 5D Planck mass,  $\gamma^{MN}=\frac{1}{2}[\gamma^M,\gamma^N]$ for the 5D gamma matrices
  $\gamma^M=(\gamma^\mu, \gamma_5)$, and $k_{\rm CS}$ is the integer-valued Chern-Simons coefficient.  For simplicity,  we limit the discussion to the compactification on
a flat orbifold $S^1/Z_2$ with radius $R$, which is described by
the 5-th coordinate \bea
y\cong -y \cong y+2\pi R.\eea
 The $U(1)_A$ gauge field $A_M=(A_\mu, A_5)$ obeys
 the $Z_2$-odd boundary condition (BC) giving an axion zero mode, i.e. 
 \bea
 A_\mu(x,y)=A_\mu(x,y+2\pi R) =-A_\mu(x,-y), \quad 
 A_5(x,y)=A_5(x,y+2\pi R) =A_5(x,-y),
 \eea
 while 
 $\tilde g_{MN}$ and $G^a_M$ obey the $Z_2$-even BC giving  the 4D graviton and gluon zero modes. We also impose  $Q_I(y)=Q_I (y+2\pi R)=\gamma_5 Q_I(-y)$, which would give 4D chiral fermion zero modes.

 The axion zero mode might be defined as  
 \bea
\frac{a(x)}{f_a}\equiv \oint dy\,   A_5(x,y), \label{5d_axion}
 \eea
for which the axion periodicity  $a(x)\cong a(x)+2\pi f_a$ is assured  by the $U(1)_A$ gauge transformation
$A_5\rightarrow A_5 +\frac{1}{R}$ . Note that a generic $U(1)_A$ gauge transformation is defined as
 \bea
U(1)_{A}: \,\,\,  A_M\rightarrow A_M+\partial_M\Lambda
\eea
for $\Lambda$ obeying the BC:
\bea \Lambda(x,y)=\Lambda(x, y+2\pi R) =-\Lambda(x,-y)\,\,\, \mbox{\emph{mod\,  $2\pi$}},\eea 
and  $\Lambda= y/R$
 is a genuine gauge transformation on $S^1/Z_2$. 
Note also that the non-linear PQ symmetry $a(x)/f_a\rightarrow a(x)/f_a+\alpha$  \,$(\alpha=\mbox{real constant})$ in the GKR basis can be identified as a
\emph{locally well-defined, but globally ill-defined} $U(1)_A$ transformation in the limit when all $U(1)_A$-charged fields are integrated out, i.e.
 \bea
 U(1)_{\rm PQ}:\,\,\, A_5\,\,\rightarrow \,\, A_5+\partial_y\tilde\Lambda 
 \,\,\,\,\mbox{for} \,\,\,\tilde\Lambda= \frac{\alpha y}{2\pi R}. \label{5d_pq}
 \eea
This implies that $U(1)_{\rm PQ}$ can be broken {\it only} by non-local effects associated with $U(1)_A$-charged field $\Phi_C$ on $S^1/Z_2$, which would be
  exponentially suppressed as $e^{-2\pi M_{\Phi_C} R}$ in the limit $R\gg 1/M_{\Phi_C}$.

To examine the low energy couplings of the axion (\ref{5d_axion}),
one may perform the dimensional reduction of the model.
 For the zero mode fluctuations given by
\bea
\hskip -1cm
\tilde g_{\mu\nu}=g_{\mu\nu}(x),   \quad G^a_\mu = G^a_\mu(x),  \quad A_5 = \frac{1}{2\pi R}\frac{a(x)}{f_a}, \quad
\big(Q_i, Q_i^c\big) = \frac{(q_i(x), q_i^c(x))}{\sqrt{\pi R}},
\eea
one finds  the 4D effective Lagrangian
\bea
{\cal L}_{4D}&=&\frac{1}{2}M_P^2 R(g) -\frac{1}{4g_{s}^2}G^{a\mu\nu}G^a_{\mu\nu}+\frac{1}{2}\partial_\mu a\partial^\mu a+\sum_{\psi=q_i,q^c_i}i\bar \psi \gamma^\mu D_\mu \psi  \nonumber \\
&+& \frac{c_G}{32\pi^2} \frac{a(x)}{f_a} G^{a\mu\nu}\tilde G^a_{\mu\nu} + \sum_{\psi=q_i,q^c_i}
c_{\psi} \frac{\partial_\mu a}{f_a} \bar \psi \gamma^\mu \psi, 
\eea
where  $M_P^2 =\pi R M_5^3$ and  $g_s^2={g_{5S}^2}/{\pi R}$ denote the 4D Planck scale and the 4D $SU(3)_c$ gauge coupling, respectively, and the axion scale and couplings are determined as 
\bea
  f^2_a =\frac{1}{4\pi g_{5A}^2 R},\quad c_G=k_{\rm CS}, \quad  c_\psi=  \frac{\mu_{Q,Q^c}}{\pi R}. \label{5d_4d}
\eea


{
It has been conjectured in \cite{wgc} that for  a $D$-dimensional $U(1)$ gauge field compatible with quantum gravity,  there should exist a $U(1)$-charged particle with a mass obeying  $m \lesssim g_D/G_D^{1/2}$ (there can be  a coefficient of  ${\cal O}(1)$ in this upper bound, which will be ignored), where
 $g_D$ and $G_D$  are the $D$-dimensional $U(1)$ gauge coupling and the $D$-dimensional Newton's constant, respectively.
This conjecture goes under the name of \textit{Weak Gravity Conjecture} (WGC), 
and can be extended to general $p$-forms in $D$ dimensions \cite{Harlow:2022gzl}. 
Axions can be seen as $0$-form gauge fields, and the WGC can be formulated using the following analogy: the gauge coupling $g_D$ becomes the inverse decay constant $1/f_a$, and the charged object is an instanton with
the Euclidean action  $S_{\text{inst}}$ which is the analogue of the mass. Then the axion WGC states that there must exist an instanton satisfying
		$S_{\text{inst}} \lesssim M_P/f_a,$ where a coefficient of  $\mathcal{O}(1)$ is again ignored.

Applying the above WGC to our case,
the UV completion of  the 5D model  (\ref{5d_model})  should include  $U(1)_A$-charged matter field
$\Phi_C$
with a mass  \bea
M_{\Phi_C}\lesssim g_{5A}M_5^{3/2}.\label {5d_wgc}\eea
 In the limit $M_{\Phi_C}\gg 1/R$ which we are concerned with, 
the Euclidean worldline of $\Phi_C$ winding the covering space of $S^1/Z_2$ can be interpreted as a worldline instanton ($\equiv$ the $\Phi_C$-instanton) with the Euclidean action \bea
S_{\rm inst}=2\pi R M_{\Phi_C}.\eea
For this instanton action and the axion decay constant $f_a$ obtained from dimensional reduction as (\ref{5d_4d}), 
 the 5D WGC  (\ref{5d_wgc}) leads to
\bea
\frac{f_a}{M_P}
\lesssim \frac{1}{S_{\rm inst}}. 
\eea
This indicates that  the $\Phi_C$-instanton corresponds to 
 the instanton required by the  axion WGC for the 5D model 
(\ref{5d_model}). One also finds that the quantum fluctuations of $\Phi_c$ generate an axion potential \cite{5d_axion1} 
\bea
\delta V_{\rm axion}\sim \frac{1}{(\pi R)^4}e^{-S_{\rm inst}}\cos\Big(\frac{a}{f_a}+\delta\Big), \label{wgc_axion}
\eea
which can be interpreted as a potential generated by
the $\Phi_C$-instanton \cite{Reece:2023czb}. Note that generically $\delta={\cal O}(1)$ for $\langle a\rangle/f_a$ identified as
 $\bar\theta$  in the SM (see (\ref{repara})), therefore $\delta V_{\rm axion}$ can spoil the PQ solution
 of the strong CP problem \emph{unless} $\delta V_{\rm axion}< 10^{-10}f_\pi^2 m_\pi^2$.}

 As the Chern-Simons coefficient $k_{\rm CS}$ is an integer, 
the axion coupling $c_G$ to 4D gluons is integer-valued as required. 
The axion couplings to 4D chiral fermions  are given by
\bea c_{\psi}=  \frac{2\mu_{Q,Q^c}M_{\Phi_C}}{S_{\rm inst}}\sim \frac{1}{S_{\rm inst}},
\eea
where we assumed\footnote{For  the 5D model
under discussion, this may look like  an ad hoc assumption. Yet it applies for  the parameters describing the  axions from $p$-form gauge fields  in string theory.} 
$\mu_{Q,Q^c}M_{\Phi_C}={\cal O}(1)$.  Then, adding the leptons to the model (\ref{5d_model}),  the axion couplings to the light quarks (at $\mu\sim 1$ GeV) and electron (at $\mu=m_e$) are estimated as 
\bea
C_{u,d,e}\sim \frac{1}{S_{\rm inst}}. \label{5d_axion_coupling}\eea

What would be the probable value of the instanton action $S_{\rm ins}$? For a QCD axion, to solve the strong CP problem, one needs
$\delta V_{\rm axion} < 10^{-10} m_\pi^2 f_\pi^2$, 
which requires 
$S_{\rm ins}\gtrsim 60-180$ for $1/R=10^3-10^{16}$ GeV. For ultralight ALP, one needs $\delta V_{\rm axion} \lesssim m_a^2f_a^2$, giving a numerically similar lower
 bound on $S_{\rm inst}$. On the other hand,
if $S_{\rm inst}=2\pi R M_{\Phi_C}\gg 10^2$, 
the 4D QCD coupling $g_s^2=g_{5S}^2/\pi R$ at $\mu\sim 1/R$  would be too small to be phenomenologically viable for reasonable values of
 $g_{5S}^2$ and $M_{\Phi_C}$.  These  imply that  $S_{\rm inst}$ needs to have a value 
 of  ${\cal O}(10^2)$.

 \subsubsection{Axions from $p$-form gauge fields in string theory}\label{sec_axionspforms}
 
 String/M theory involves a variety of extended objects, i.e. $(p-1)$-dimensional branes, 
 which couple to  $p$-form gauge fields $A_p$
 with the associated $(p-1)$-form gauge symmetry \cite{ibanez}: \bea
 G_{p-1}:\quad A_p\rightarrow A_p+d\Lambda_{p-1} .\label{eq_gaugesymm}\eea
 Upon compactification, the zero modes of $A_p$ can be identified as 4D axions whose periodicity is assured by the
 quantized  charges of $G_{p-1}$.
 For  compactifications involving the $p$-cycles $\Sigma_p^{(i)}$ ($i=1,.., N_p$)  in the internal space,  the axion zero modes\footnote{For 2-form gauge field $A_2$,  there is an additional axion zero mode given by
$\partial_\mu \theta(x) = \epsilon_{\mu\nu\rho\sigma}\partial^\nu A_2^{\rho\sigma}(x)$, which is often called the model-independent axion \cite{string_axion1}.} 
 are given by \cite{string_axion1,string_axion2}
 \bea
 A_p(x,y) =\sum_i 
\theta_i(x)\omega_p^{(i)}(y),
  \label{string_axion}
 \eea
 where 
$\theta_i(x)\equiv {a_i(x)}/{f_i}\cong \theta_i(x)+2\pi$, and $\omega_p^{(i)}$ are the harmonic $p$-forms dual to $\Sigma_p^{(i)}$, i.e. $\int_{\Sigma_p^{(j)}}\omega_p^{(i)}=\delta^i_j$.
Then the PQ transformation $a_i/f_a\rightarrow a_i/f_i+\alpha_i$ $(\alpha_i=\mbox{constant})$
corresponds to
\bea
U(1)_{\rm PQ}^{(i)}:\quad 
A_p \rightarrow A_p +\alpha_i \omega_p^{(i)}.\eea
As $\omega_p^{(i)}$ is locally an exact $p$-form, but {not globally}, $U(1)_{\rm PQ}^{(i)}$  can be identified as a locally well-defined, but globally ill-defined  $G_{p-1}$ gauge transformation.

For compactifications preserving 4D $\mc{N}=1$ SUSY,
the low energy properties of  the axion zero modes (\ref{string_axion})   can be described by the 4D $\mc{N}=1$  supergravity (SUGRA) Lagrangian. 
In such compactifications, for each axion $\theta_i=a_i/f_i$, there exists a saxion (modulus) partner $\tau_i$, forming the  scalar component of chiral superfield (see Sec.\ref{sec:stringaxion_IIB} for more details) as
 \bea
 T_i = \tau_i +i\theta_i.\eea
Also, for $\theta_i\cong \theta_i+2\pi$,  the vacuum value of $\tau_i$ 
 can be  identified as the Euclidean action of the brane instanton\footnote{
 For the model-independent axion $\partial_\mu \theta(x) = \epsilon_{\mu\nu\rho\sigma}\partial^\nu A_2^{\rho\sigma}(x)$, the associated instanton is 
the Euclidean 5-brane  wrapping the 6D internal space,
which magnetically couples to $A_2$ \cite{ibanez}.}
which corresponds to  the Euclidean $(p-1)$-brane wrapping $\Sigma_p^{(i)}$, i.e.
\bea
\langle \tau_i\rangle =S_{\rm inst}^{(i)} \propto \mbox{Vol}(\Sigma_p^{(i)}). \label{brane_action}\eea 

 In 4D $\mc{N}=1$ SUGRA, the axion couplings and scales are determined by
the
 K\"ahler potential $K$ and the holomorphic gauge kinetic function ${\cal F}_A$ for which
  $\langle {\rm Re}({\cal F}_A) \rangle={1}/{g_A^2}$.
Keeping only the relevant terms,  $K$ and ${\cal F}_A$ take the form
 \bea
 8\pi^2{\cal F}_A = c_{Ai} T^i, \quad 
 K= K_0(T^i+T^{i*}) + Z_\Phi(T^i+T^{i*})\Phi^*\Phi,
 \eea
 where $\Phi$ stands for gauge-charged matter fields with the K\"ahler metric $Z_\Phi$, and
 $c_{iA}$ are integers for properly normalized gauge couplings $g_A^2$.
 The resulting 4D
 Lagrangian \cite{nilles} includes   
 \bea
  \frac{1}{2} (f_a^2)_{ij}\partial_\mu \theta_i \partial^\mu\theta_j  
 +\frac{c_{iA}}{32\pi^2} \theta_i F^{A\mu\nu} {\tilde F}^A_{\mu\nu} 
+  \partial_\mu \theta_i \Big[ic_{i\phi}\big(\phi^*D_\mu\phi -{\rm h.c.}\big)  + c_{i\psi} \bar\psi\bar\sigma^\mu\psi\Big],
\label{4d_axion}\eea
where $\phi$ and $\psi$ are the {canonically normalized} scalar and fermion components of $\Phi$, and
\bea 
(f_a^2)_{ij} = 2M_P^2\frac{\partial^2 K_0}{\partial T^i\partial T^{j*}},\quad
 c_{iA}=8\pi^2 \frac{{\cal F}_A}{\partial T^i}, \quad
c_{i\phi} = \frac{\partial \ln Z_\Phi}{\partial T^i}, \quad  c_{i\psi}=\frac{\partial \ln (e^{-K_0/2}Z_\Phi)}{\partial T^i}. \label{sugra_coupling}\eea

 Generically the brane instanton  with the Euclidean action (\ref{brane_action}) 
 can give a non-perturbative correction to
 the superpotential or to the K\"ahler potential \cite{string_instanton1,string_instanton2}, yielding
 (see Sec.\ref{sec:stringaxion_potential})
\bea\delta V_{\rm axion}=M_i^4 e^{-\langle \tau_i\rangle}\cos(\theta_i+\delta_i)\quad \big(M_i^4\sim m_{3/2}M_P^3
\,\,\, \mbox{or}\,\,\, m_{3/2}^2M_P^2\big).
\eea One may now repeat the argument  to estimate $S_{\rm inst}$
 for the axion from 5D gauge field (see the discussion below (\ref{5d_axion_coupling})). It again results in 
\bea
S_{\rm inst}=\langle \tau\rangle ={\cal O}(10^2)\label{braneinstanton_action}\eea
for the brane instanton associated with
the QCD axion or ultralight ALP  from  a $p$-form gauge field in string theory.

One can now extract  some qualitative feature of the axion scale and couplings.
For  relatively simple compactifications  \emph{not} generating a big scale hierarchy, (\ref{sugra_coupling}) and (\ref{braneinstanton_action}) imply
\bea
f_a \sim \frac{M_P}{S_{\rm inst}}\sim 10^{16}\, {\rm GeV}, \quad c_\phi\sim c_\psi \sim \frac{1}{S_{\rm inst}}\sim 10^{-2}.\eea
On the other hand,
for more involved compactifications generating  either a large  volume \cite{Balasubramanian:2005zx} or a strong warping
\cite{rs}, $f_a$ can be lowered 
 by the
large volume factor \cite{Cicoli:2012sz,Cicoli:2021gss} or red-shifted by an exponentially  small warp factor \cite{5d_axion2}.
As $c_{\phi,\psi}$ are not significantly affected by this rescaling of $f_a$,
the  QCD 
axion or ultralight ALPs 
from $p$-form gauge fields  in string theory  can in principle have  $f_a$ \emph{anywhere} in the range
${\cal O}(10^8-10^{16})$  {GeV}, while
their low energy couplings to the light quarks and electron are estimated as
\bea
C_{u,d,e}\sim \frac{1}{S_{\rm inst}}={\cal O}(10^{-2}).\eea

 \subsection{Discriminating between different axion models with low energy observables}

In the previous section, we discussed  a variety of  models in which a light axion arises  from 
either the spontaneous breakdown of a linear $U(1)_{\rm PQ}$ or a higher-dimensional gauge field.
For the purpose of presentation, let us  call the axion from linear $U(1)_{\rm PQ}$ ``field-theoretic axion'' and the axion from higher-dimensional gauge field  ``string-theoretic  axion''\footnote{This distinction is interesting in the context of quantum gravity. For instance,  $f_a$ for a field-theoretic axion vanishes at the origin in field space, while
$f_a\rightarrow 0$  for a string-theoretic axion corresponds to an infinitely distant point where the 4D effective theory breaks down \cite{Ooguri:2006in,Reece:2023czb}. In particular, the UV cutoff can be estimated as $\Lambda_{\text{UV}}\lesssim \sqrt{f_a \mpl} $ \cite{Reece:2018zvv}. 
}, although it should be noted that string theory can also provide  field-theoretic axions\footnote{For instance, in string compactifications with nonzero $U(1)_X$ magnetic flux, some of the string-theoretic axions can transform non-linearly under the $U(1)_X$ gauge symmetry. Then
a combination  of  $U(1)_X$-charged string-theoretic axions is eaten by the $U(1)_X$ gauge boson while leaving
a \emph{linear} $U(1)_{\rm PQ}$ which is the global $U(1)_X$ transformation applied \emph{only} for matter fields, \emph{not} for the eaten string-theoretic axion \cite{string_axion1,lukas,choi_jeong}.  This linear $U(1)_{\rm PQ}$ can be spontaneously broken  at lower energy scale by the vacuum value of some $U(1)_X$-charged matter field, thereby giving a field-theoretic axion.}.

In view of their motivation and the prospect for experimental detection, there are two kinds of particularly interesting  axions, the ``QCD axion'' solving the strong CP problem with a nonzero coupling to the gluons  ($c_G\neq 0$), and  the ``ultralight (UL) ALP'' with  nonzero coupling to the photon
 ($c_G=0, \, c_\gamma=c_{W}+c_B \neq 0$).  Let us examine to what extent
we can discriminate between different models for QCD axion or UL ALP with experimentally measured axion mass and couplings.
Considering  only relatively simple models with  $c_{G,W,B}={\cal O}(1)$,
we first find 
\bea
\mbox{QCD axion:}\,\,\,\, \frac{m_a}{g_{a\gamma}}\sim 10 \,\,  {\rm GeV}^2,\quad \mbox{UL ALP:} \,\,\,\, \frac{m_a}{g_{a\gamma}}\ll 10 \,\,  {\rm GeV}^2.\eea
For further discrimination, we can also examine the coupling ratios  $g_{aX}/g_{a\gamma}$ ($X=p,n,e$).
 From the results in Sec.\ref{sec:axion_eft} and Sec.\ref{sec:model}, we then obtain the following order of magnitude estimates: 
\bea
&& \hskip -1.8cm \mbox{* \, DFSZ QCD axion}: \quad 
\frac{g_{ap}}{g_{a\gamma}}\sim  \frac{g_{an}}{g_{a\gamma}}\sim\frac{g_{ae}}{g_{a\gamma}} \sim 10^{3}, \nonumber \\
&& \hskip -1.8cm \mbox{* \, KSVZ (or composite) QCD axion:}\quad 
\frac{g_{ap}}{g_{a\gamma}}\sim  20\,\frac{g_{an}}{g_{a\gamma}}\sim 10^3\frac{g_{ae}}{g_{a\gamma}} \sim 10^{3},\nonumber \\
&& \hskip -1.8cm \mbox{* \, String theoretic QCD axion:}\quad 
\frac{g_{ap}}{g_{a\gamma}}\sim  20\,\frac{g_{an}}{g_{a\gamma}}\sim 10^2\frac{g_{ae}}{g_{a\gamma}} \sim 10^{3}, \nonumber \\
&&  \hskip -1.8cm\mbox{* \, KSVZ (or composite) UL ALP:}\quad 
\frac{g_{ap}}{g_{a\gamma}}\sim  \,\frac{g_{an}}{g_{a\gamma}}\sim (1-10)\times\frac{g_{ae}}{g_{a\gamma}} \sim 10^{-1}-1, \nonumber \\
&& \hskip -1.8cm \mbox{* \, String theoretic UL ALP:}\quad 
\frac{g_{ap}}{g_{a\gamma}}\sim  \frac{g_{an}}{g_{a\gamma}}\sim \frac{g_{ae}}{g_{a\gamma}} \sim 10. \label{5axions}
\eea
The above results imply that we might be able to discriminate between different axion models  with experimentally measured axion mass and axion couplings. In particular, measuring $m_a, g_{a\gamma}$ and $g_{ae}$ may allow us to discriminate between all five different axions listed in (\ref{5axions}).
\section{Model building for string axions} \label{sec_stringaxion}
In this section, we show how to derive the effective theory of axions from string theory. Specifically, we will work in type IIB superstring theory compactified on Calabi-Yau orientifolds with O3/O7 planes. Although all the five 10D superstring theories could be used, type IIB has proven to be particularly suitable for model building. Our focus will be on those axions coming from the reduction of $C_4$ gauge potentials, i.e. for the sake of exposition we are considering orientifolds projecting out part of the axion spectrum. A large portion of the discussion also applies to the axions deriving from $B_2$ and $C_2$ forms on 2-cycles, however for a more detailed treatment we refer the reader to refs. \cite{Cicoli:2021tzt,Cicoli:2021gss,Grimm:2007xm,McAllister:2008hb}.

\subsection{Axions in type IIB string theory} \label{sec:stringaxion_IIB}

We consider 10D type IIB string theory  with $\mc{N}=2$ supersymmetry (32 supercharges) on a manifold with metric $G_{MN}$. The action in Einstein frame reads
 \begin{equation}
     S_{\text{IIB}}=\frac{1}{2\kappa_{10}^2}\int d^{10}x \sqrt{-G}\left({\cal R}_{10}-\frac{\partial_M \tau\partial^M\bar{\tau}}{2(\im \tau)^2}-\frac{\left|G_3\right|^2}{12 \im\tau}-\frac{|F_5|^2}{4\cdot 5!}\right)+ \frac{1}{8 i \kappa_{10}^2} \int  \frac{C_4\wedge G_3\wedge \bar{G}_3}{\im \tau}\coma
 \end{equation}
where $G_3=F_3-\tau H_3$, $\tau= C_0+i e^{-\phi}$ is the axio-dilaton and $\kappa_{10}^2\sim l_s^8$ is the 10D gravitational coupling, with $l_s^8=2\pi\sqrt{\alphap}$, $l_s$ being the string length. The field strengths are defined in terms of the gauge potentials as
\begin{equation}
        H_3=dB_2\coma\, F_3=dC_2\coma\, \tilde{F}_5=dC_4-\frac{1}{2}C_2\wedge dB_2+\frac{1}{2}B_2\wedge dC_2\fstop
\end{equation}
 Given that we are interested in the phenomenology, we have to first lower the number of dimensions down to four. This translates into finding a solution for the 10D equations of motion with non-trivial Riemann tensor, that nevertheless solve the vacuum Einstein's
 equations $R_{MN}=0$, i.e. the extra dimensions must be described by a Ricci-flat manifold. A non-trivial class of such manifolds is called Calabi-Yau (CY) threefold. We proceed with the ansatz of a 10D product manifold $\mc{M}_{10}=\mathbb{R}^{1,3}\times \text{CY}_3$ and compactify on the latter. This produces two important outcomes: (i) the theory is now 4D with 8 supercharges and (ii) we have a number of massless scalars counted by two topological quantities of the CY, the Hodge numbers $h^{2,1}$ and $h^{1,1}$. The first counts the number of \emph{complex structure moduli} which describe the shape of the CY, while $h^{1,1}$ gives the number of \emph{K\"ahler moduli} parametrizing its size. These numbers range from order 1 to order few hundreds \cite{Candelas:1987kf,Kreuzer:2000xy}.

To reduce further the amount of supersymmetry and arrive as close as possible to describe our universe while maintaining computational control, we can remove half of the supercharges by incorporating orientifold planes. Since these objects carry negative tension and negative charge with respect to the gauge potentials, they also balance the positive charges sourced by fluxes and branes. Orientifold planes project out half of the spectrum and divide the Hodge numbers into even and odd under the involution, $h^{p,q}_{\pm}$. Hence, the spectrum in our final 4D, $\mc{N}=1$ (i.e. 4 supercharges) low energy theory is given by
\begin{equation}
    \begin{split}
        &T_i \;\, \text{K\"ahler moduli,}\; i=1,\dots,h^{1,1}_+\coma\quad\quad G_\alpha \;\,\text{2-form axions,}\;\alpha=1,\dots,h^{1,1}_-\coma\\
         &V_m \;\, \text{vector multiplets,}\; m=1,\dots,h^{2,1}_+\coma\, U_a \;\, \text{complex str. moduli,}\; a=1,\dots,h^{2,1}_-\coma 
    \end{split}
\end{equation}
together with the axio-dilaton. In addition to the real part of the axio-dilaton $C_0$, the (closed string) axions $\theta_i$, $c_\alpha$, $b_\alpha$ are encoded in the complex fields $T_i$ and $G_\alpha$ as \cite{Grimm:2004uq,Grimm:2007xm} 
\begin{equation}
\begin{split}
        T_i= \tau_i +i(\theta_i -\frac{1}{2}\kappa_{i \alpha\beta}c^\alpha b^\beta)+\frac{\kappa_{i\alpha\beta}}{2(\tau-\bar{\tau})}G^\alpha(G^\beta-\bar{G}^\beta)\coma\; 
        G_\alpha=c_\alpha-\tau b_\alpha\coma
\end{split}  
\end{equation}
where $\kappa_{i\alpha\beta}$ are topological numbers of the CY and the $\tau_i=\frac{1}{2}\kappa_{ijk}t^jt^k$ are the volumes of the 4-cycles $\Sigma_4^{(i)}$ expressed as functions of the 2-cycle volumes $t_i$. The axion fields $b_\alpha$, $c_\alpha$, $\theta_i$ arise, respectively, from the integration of the 2-forms $B_2$ and $C_2$ over 2-cycles $\Sigma_2^{(\alpha)}$, and from the integration of the 4-form $C_4$ over 4-cycles $\Sigma_4^{(i)}$, namely
\begin{equation}
    b_\alpha=\frac{1}{l_s^2}\int_{\Sigma_2^{(\alpha)}}B_2\coma \,c_\alpha=\frac{1}{l_s^2}\int_{\Sigma_2^{(\alpha)}}C_2\coma\, \theta_i=\frac{1}{l_s^4}\int_{\Sigma_4^{(i)}}C_4\fstop 
\end{equation}
Because of their origin, in the low energy theory they enjoy a continuous shift symmetry inherited from the higher dimensional $p$-form gauge symmetry, as explained in Sec.\ref{sec_axionspforms}. For simplicity, in what follows we focus on orientifolds with $h^{1,1}_- = h^{2,1}_+= 0$ and hence on the axions $\theta_i$.

\subsection{Axion potential}\label{sec:stringaxion_potential}
After compactifying on a CY threefold with orientifolds, the 4D effective theory contains many massless scalar fields. In the following, we consider a setup in which the axio-dilaton and the complex structure moduli are stabilized at high energies by fluxes \cite{Giddings:2001yu}. 
At tree-level in $g_s$ and $\alphap$, the K\"ahler moduli are massless and uncharged scalar fields which, thanks to their effective gravitational coupling to all SM particles, would mediate undetected long-range fifth forces and affect the Big Bang nucleosynthesis. Moreover, if these fields were to be massless during inflation, they could spoil the slow-roll regime. This \emph{cosmological moduli problem} can be avoided by generating a potential for these particles and hence give them a mass at energies above the Big Bang nucleosynthesis one.

The low-energy theory is a SUGRA theory. The (F-term) $4$D scalar potential is given in terms of a K\"ahler potential $K$ and a superpotential $W$ as
\begin{equation}\label{eq_VSUGRA}
    V=e^{K}\left[K^{i\bar{j}}\mathcal{D}_i W\mathcal{D}_{\bar{j}}\overline{W}-3|W|^2\right]\coma
\end{equation}
where $K^{i\bar{j}}=(\partial_i\partial_{\bar{j}} K)^{-1}$ is the inverse of the K\"ahler metric, $\mathcal{D}_i W\equiv\partial_i W + K_i W$ is the K\"ahler covariant derivative and $i,j=1,\dots,h^{1,1}_+$. The minimum of $V$ preserves supersymmetry if $\mathcal{D}_{T_i} W|_{\langle T_i\rangle}=0$.
After complex structure and axio-dilaton stabilization, we can write the superpotential as $W=W_0+W_{\text{np}}(T_i)$, where $W_0$ is a constant proportional to the VEVs of $U_a$ and $\tau$ while $W_{\text{np}}$ includes the non-perturbative corrections to $T_i$. $W$ is holomorphic and receives no perturbative corrections. $W_{\text{np}}$ can be generated either by Euclidean D$3$-brane instantons or by gaugino condensation on stacks of D$7$-branes wrapping 4-cycles. Both these contributions read
\begin{equation}\label{eq:npsuperpot}
 	W_{\text{np}}= \sum_i A_i e^{-S_{\text{inst}}}= \sum_i A_i e^{-\mathfrak{a}_i T_i}\coma
 \end{equation}
 where $\mathfrak{a}_i =2\pi$ for ED$3$-branes and $\mathfrak{a}_i=2\pi/c(G_i)$ for the gaugino condensation case, $c(G_i)$ being the dual Coxeter number of the gauge group $G_i$ on the $i$-th stack of D$7$-branes. The 1-loop Pfaffians $A_i$ depend on the stabilization of $U_a$ and $\tau$. 
The K\"ahler potential is not holomorphic and can receive both perturbative and non-perturbative corrections such that generically $K= K_{\text{tree}}+K_{\text{p}}+K_{\text{np}}$. The tree-level piece is given by $K_{\text{tree}}=-2\log(\mc{V})$, where $\mc{V}=\frac{1}{6}\kappa_{ijk}t^it^jt^k$ is the overall volume of the CY. Note that it is not always possible to write $\mc{V}$ in terms of the K\"ahler coordinates $T_i$, so the dependence of $K$ on $\tau_i$ is often implicit. Let us include also the leading perturbative corrections, such that 
\begin{equation}\label{eq_kpotlvs}
   K=-2\log(\mc{V}+\hat\xi/2)\coma
\end{equation} 
where $\hat\xi$ is proportional to $g_s^{-3/2}$ \cite{Becker:2002nn}.   

Quantum corrections (as the ones mentioned above) are crucial to stabilize the remaining massless moduli. The way in which those corrections are present and their magnitude result in different stabilization regimes. The most studied approaches are KKLT \cite{kklt} and the Large Volume Scenario \cite{Balasubramanian:2005zx}, see \cite{McAllister:2023vgy} for a recent review. Schematically, the former does not need corrections to $K$ but requires $|W_{\text{np}}|\sim |W_0|$, such that $W_0$ has to be tuned small. LVS instead relies on the presence of $K_{\text{p}}$ as in (\ref{eq_kpotlvs}) and needs the CY volume to be large.

Finally, plugging the quantum-corrected $K$ and $W$ in (\ref{eq_VSUGRA}), the axion-dependent part of the total potential is given by \cite{Conlon:2006tq}
\begin{equation}\label{eq_axionpot}
    \begin{split}
    V_{\text{axion}}=e^K\bigg(&K^{i\bar{j}}\Big(2 \mathfrak{a}_i\mathfrak{a}_j|A_iA_j|e^{-\mathfrak{a}_i\tau_i-\mathfrak{a}_j\tau_j}\cos(\mathfrak{a}_i\theta_i+\mathfrak{a}_j\theta_j+\gamma_{ij})\Big)\\
    &- 4\mathfrak{a}_i\tau_i |W_0 A_i|e^{-\mathfrak{a}_i\tau_i}\cos(\mathfrak{a}_i\theta_i+\beta_i)- 4\mathfrak{a}_i\tau_i |A_iA_j|\cos(\mathfrak{a}_i\theta_i+\mathfrak{a}_j\theta_j+\gamma_{ij})
  \bigg)\coma
    \end{split}
\end{equation}
where $\gamma_{ij}$ and $\beta_i$ are phases. Now, we can compute the mass matrix as $M_{ij}^2=\partial_{\theta_i}\partial_{\theta_j}V_{\text{axion}}$, and the eigenvalues of this matrix correspond to the mass-squared of the canonically normalized axions. How to go from string-theoretical to canonically normalized axions is the subject of what follows.  

\subsection{Kinetic terms and canonical normalization}
The kinetic terms of the effective Lagrangian are completely specified by the K\"ahler metric $K_{i\bar{j}}=\partial_i\partial_{\bar{j}} K$.
At the perturbative level the 10D gauge invariances of the $p$-form gauge fields of (\ref{eq_gaugesymm}) descend to continuous shift symmetries of their associated axions in 4D: $\theta_i\sim \theta_i+c$, $c\in \mathbb{R}$. The Lagrangian for massless axions reads 
\begin{equation}
    \mc{L}\supset \frac{\partial^2 K}{\partial T^i \partial \bar{T}^j} \partial_\mu \theta^i\partial^\mu \theta^j\fstop
\end{equation}
In order to work with canonically normalized fields, we need to diagonalize the K\"ahler metric and find the eigenvalues $\lambda_i$ and eigenvectors $\tilde{\theta}_i$. We define the canonically normalized axion fields as $a_i =\sqrt{\lambda_i}\, \tilde{\theta}_i $ such that 
\begin{equation}
	\mc{L}_{\text{kin}}\supset \frac{\lambda_i}{2} \partial_\mu \tilde{\theta}_i \partial^\mu \tilde{\theta}_i=
	\frac12 \partial_\mu a_i \partial^\mu a_i \fstop
\end{equation}
In the case of massless axions, it is  common to refer to $\hat{f}_i = \sqrt{\lambda_i}$ as the axion decay constant, because the couplings of the physical axions with all other fields scale as $1/\hat{f}$, cf. (\ref{sugra_coupling}). So far we have only considered massless axions but, as with the rest of the moduli, these fields need to be stabilized. Axions acquire a mass through the non-perturbative quantum corrections in (\ref{eq:npsuperpot}) that break the continuous shift symmetry down to its discrete subgroup. The typical form of the potential arising from a single non-perturbative correction reads (cf. (\ref{eq_axionpot}))
\begin{equation}
	V(a_i)=\Lambda_i^4 \cos(\mathfrak{a}_i \theta_i)\coma
\end{equation}
where $\Lambda_i$ is a dynamically-generated scale proportional to $e^{-S_{\rm inst}}$, cf. Sec.\ref{sec_axionspforms}. To work with physical fields, we need to find a basis that diagonalizes both the mass matrix and the field space metric. Note that this is not always possible, and in general one is able to diagonalize only either the kinetic terms or the potential. In the simplest case where the K\"ahler metric is approximately diagonal ($\theta_i\sim \tilde{\theta}_i$) and we have a single non-perturbative correction, computing the decay constant becomes rather simple. Since the field periodicity corresponds to that of the potential, using $a_i =\sqrt{\lambda_i}\, \theta_i $, the axion decay constant $f_i$ derives from \cite{Cicoli:2021gss}:
\begin{equation}
 \mathfrak{a}_i \theta_i\rightarrow  \mathfrak{a}_i\theta_i+ 2\pi k \quad\Rightarrow
		\quad a_i\rightarrow a_i + 2\pi k f_i \coma
\mbox{where}\quad f_i =\frac{\sqrt{\lambda_i}}{\mathfrak{a}_i}\mpl\fstop
\end{equation}


\subsection{String axions as dark matter}
The phenomenology of string axions is characterized by their masses and decay constants. From the discussion above, we see that generically they scale as
\begin{equation}
    \frac{m_i^2}{\mpl^2}\sim \frac{\mathfrak{a_i}\tau_i|W_0|}{\mc{V}^2}e^{-\mathfrak{a}_i\tau_i}\coma \frac{f_i}{\mpl}\sim\frac{1}{\mathfrak{a}_i\tau_i}\fstop
\end{equation}
For example, we can use these quantities to compute the abundance of dark matter when composed by UL ALPs as~\cite{Cicoli:2012sz}:
\begin{equation}
	\label{eq:DMabundance}
	\frac{\Omega_{a}h^2}{0.112}\simeq 2.2 \times \left(\frac{m_a}{10^{-22} \mbox{ eV}}\right)^{1/2}\left(\frac{f_a}{10^{17}\mbox{ GeV}}\right)^2 \theta_{\text{m}}^2\coma
\end{equation}
where $\theta_{\text{m}}\in [0,2\pi]$ is the initial misalignment angle. Eq. (\ref{eq:DMabundance}) holds for $f_a$ larger than the inflationary scale and for $m \gtrsim 10^{-28}\text{ eV}$, i.e. to axions which oscillate before matter-radiation equality. By considering different setups (where the moduli are appropriately stabilized) and plugging the resulting values of $m_a$ and $f_a$ in (\ref{eq:DMabundance}), ref. \cite{Cicoli:2021gss} predicted the parameter space spanned by different types of string axions. While a portion of this space is already excluded by experimental constraints, interestingly a larger part will be covered by future searches. Hence, if at some point axions were to be found, we may be able to learn from the data about the type of axion detected, its couplings and potentially even something about their underlying microscopic theory.



\section{Conclusion}
Axions have been postulated with many motivations in particle physics and cosmology. 
In regard to their UV origin, there are two types of axions,  a field-theoretic axion arising from the spontaneous breakdown of
a linearly realized Peccei-Quinn $U(1)$ symmetry in the 4D theory, and  a string-theoretic axion originating from  a gauge field in the higher dimensional theory.    
 In this lecture, we discussed some
theory and model building aspects of axion physics for both types of axions, in particular the possible connection between the UV origin of axions and the associated low energy axion physics. We also gave an introduction to the effective theory of string-theoretic axions in the latter part of the lecture.
 Interestingly,  different axion models predict distinctive pattern of low energy axion couplings, which might be
 testable in future axion detection experiments and also have interesting  implications for astrophysical or cosmological studies of axions.

\acknowledgments
\noindent This publication is based upon work from COST Action COSMIC WISPers CA21106, supported by COST (European Cooperation in Science and Technology).
N.R. would like to thank M. Cicoli, V. Guidetti, J. Leedom, M. Putti, F. Revello and A. Westphal for discussions on the topics presented here.
K.C. is supported by IBS under the project code, IBS-R018-D1, and  N.R. is supported by a Leverhulme Trust Research Project Grant RPG-2021-423.


\bibliographystyle{JHEP}
\bibliography{refs}

\providecommand{\href}[2]{#2}\begingroup\raggedright\begin{thebibliography}{10}

\bibitem{Kim:2008hd}
J.~E. Kim and G.~Carosi {\em Rev. Mod. Phys.} {\bf 82} (2010) 557--602,
  [\href{http://arxiv.org/abs/0807.3125}{{\tt arXiv:0807.3125}}]. [Erratum:
  Rev.Mod.Phys. 91, 049902 (2019)].

\bibitem{Marsh:2015xka}
D.~J.~E. Marsh {\em Phys. Rept.} {\bf 643} (2016) 1--79,
  [\href{http://arxiv.org/abs/1510.07633}{{\tt arXiv:1510.07633}}].

\bibitem{DiLuzio:2020wdo}
L.~Di~Luzio, M.~Giannotti, E.~Nardi, and L.~Visinelli {\em Phys. Rept.} {\bf
  870} (2020) 1--117, [\href{http://arxiv.org/abs/2003.01100}{{\tt
  arXiv:2003.01100}}].

\bibitem{Choi:2020rgn}
K.~Choi, S.~H. Im, and C.~Sub~Shin {\em Ann. Rev. Nucl. Part. Sci.} {\bf 71}
  (2021) 225--252, [\href{http://arxiv.org/abs/2012.05029}{{\tt
  arXiv:2012.05029}}].

\bibitem{Hook:2018dlk}
A.~Hook {\em PoS} {\bf TASI2018} (2019) 004,
  [\href{http://arxiv.org/abs/1812.02669}{{\tt arXiv:1812.02669}}].

\bibitem{Reece:2023czb}
M.~Reece \href{http://arxiv.org/abs/2304.08512}{{\tt arXiv:2304.08512}}.

\bibitem{Ubaldi:2008nf}
L.~Ubaldi {\em Phys. Rev. D} {\bf 81} (2010) 025011,
  [\href{http://arxiv.org/abs/0811.1599}{{\tt arXiv:0811.1599}}].

\bibitem{Dine:2018glh}
M.~Dine, L.~Stephenson~Haskins, L.~Ubaldi, and D.~Xu {\em JHEP} {\bf 05} (2018)
  171, [\href{http://arxiv.org/abs/1801.03466}{{\tt arXiv:1801.03466}}].

\bibitem{Kaloper:2017fsa}
N.~Kaloper and J.~Terning {\em JHEP} {\bf 03} (2019) 032,
  [\href{http://arxiv.org/abs/1710.01740}{{\tt arXiv:1710.01740}}].

\bibitem{Peccei:1977hh}
R.~D. Peccei and H.~R. Quinn {\em Phys. Rev. Lett.} {\bf 38} (1977) 1440--1443.

\bibitem{Weinberg:1977ma}
S.~Weinberg {\em Phys. Rev. Lett.} {\bf 40} (1978) 223--226.

\bibitem{Wilczek:1977pj}
F.~Wilczek {\em Phys. Rev. Lett.} {\bf 40} (1978) 279--282.

\bibitem{Georgi:1986df}
H.~Georgi, D.~B. Kaplan, and L.~Randall {\em Phys. Lett. B} {\bf 169} (1986)
  73--78.

\bibitem{Barr:1992qq}
S.~M. Barr and D.~Seckel {\em Phys. Rev. D} {\bf 46} (1992) 539--549.

\bibitem{Kamionkowski:1992mf}
M.~Kamionkowski and J.~March-Russell {\em Phys. Lett. B} {\bf 282} (1992)
  137--141, [\href{http://arxiv.org/abs/hep-th/9202003}{{\tt hep-th/9202003}}].

\bibitem{Holman:1992us}
R.~Holman, S.~D.~H. Hsu, T.~W. Kephart, E.~W. Kolb, R.~Watkins, and L.~M.
  Widrow {\em Phys. Lett. B} {\bf 282} (1992) 132--136,
  [\href{http://arxiv.org/abs/hep-ph/9203206}{{\tt hep-ph/9203206}}].

\bibitem{Choi:2017gpf}
K.~Choi, S.~H. Im, C.~B. Park, and S.~Yun {\em JHEP} {\bf 11} (2017) 070,
  [\href{http://arxiv.org/abs/1708.00021}{{\tt arXiv:1708.00021}}].

\bibitem{Chala:2020wvs}
M.~Chala, G.~Guedes, M.~Ramos, and J.~Santiago {\em Eur. Phys. J. C} {\bf 81}
  (2021), no.~2 181, [\href{http://arxiv.org/abs/2012.09017}{{\tt
  arXiv:2012.09017}}].

\bibitem{Bauer:2020jbp}
M.~Bauer, M.~Neubert, S.~Renner, M.~Schnubel, and A.~Thamm {\em JHEP} {\bf 04}
  (2021) 063, [\href{http://arxiv.org/abs/2012.12272}{{\tt arXiv:2012.12272}}].

\bibitem{Choi:2021kuy}
K.~Choi, S.~H. Im, H.~J. Kim, and H.~Seong {\em JHEP} {\bf 08} (2021) 058,
  [\href{http://arxiv.org/abs/2106.05816}{{\tt arXiv:2106.05816}}].

\bibitem{villadoro}
G.~Grilli~di Cortona, E.~Hardy, J.~Pardo~Vega, and G.~Villadoro {\em JHEP} {\bf
  01} (2016) 034, [\href{http://arxiv.org/abs/1511.02867}{{\tt
  arXiv:1511.02867}}].

\bibitem{chang_choi}
S.~Chang and K.~Choi {\em Phys. Lett. B} {\bf 316} (1993) 51--56,
  [\href{http://arxiv.org/abs/hep-ph/9306216}{{\tt hep-ph/9306216}}].

\bibitem{Agrawal:2017cmd}
P.~Agrawal, J.~Fan, M.~Reece, and L.-T. Wang {\em JHEP} {\bf 02} (2018) 006,
  [\href{http://arxiv.org/abs/1709.06085}{{\tt arXiv:1709.06085}}].

\bibitem{cw_axion1}
K.~Choi and S.~H. Im {\em JHEP} {\bf 01} (2016) 149,
  [\href{http://arxiv.org/abs/1511.00132}{{\tt arXiv:1511.00132}}].

\bibitem{cw_axion2}
D.~E. Kaplan and R.~Rattazzi {\em Phys. Rev. D} {\bf 93} (2016), no.~8 085007,
  [\href{http://arxiv.org/abs/1511.01827}{{\tt arXiv:1511.01827}}].

\bibitem{cw_axion3}
M.~Farina, D.~Pappadopulo, F.~Rompineve, and A.~Tesi {\em JHEP} {\bf 01} (2017)
  095, [\href{http://arxiv.org/abs/1611.09855}{{\tt arXiv:1611.09855}}].

\bibitem{ksvz1}
J.~E. Kim {\em Phys. Rev. Lett.} {\bf 43} (1979) 103.

\bibitem{ksvz2}
M.~A. Shifman, A.~I. Vainshtein, and V.~I. Zakharov {\em Nucl. Phys. B} {\bf
  166} (1980) 493--506.

\bibitem{dfsz1}
M.~Dine, W.~Fischler, and M.~Srednicki {\em Phys. Lett. B} {\bf 104} (1981)
  199--202.

\bibitem{dfsz2}
A.~R. Zhitnitsky {\em Sov. J. Nucl. Phys.} {\bf 31} (1980) 260.

\bibitem{string_axion1}
E.~Witten {\em Phys. Lett. B} {\bf 149} (1984) 351--356.

\bibitem{string_axion2}
P.~Svrcek and E.~Witten {\em JHEP} {\bf 06} (2006) 051,
  [\href{http://arxiv.org/abs/hep-th/0605206}{{\tt hep-th/0605206}}].

\bibitem{5d_axion1}
N.~Arkani-Hamed, H.-C. Cheng, P.~Creminelli, and L.~Randall {\em Phys. Rev.
  Lett.} {\bf 90} (2003) 221302,
  [\href{http://arxiv.org/abs/hep-th/0301218}{{\tt hep-th/0301218}}].

\bibitem{5d_axion2}
K.-w. Choi {\em Phys. Rev. Lett.} {\bf 92} (2004) 101602,
  [\href{http://arxiv.org/abs/hep-ph/0308024}{{\tt hep-ph/0308024}}].

\bibitem{fujikawa}
K.~Fujikawa {\em Phys. Rev. Lett.} {\bf 42} (1979) 1195--1198.

\bibitem{composite1}
J.~E. Kim {\em Phys. Rev. D} {\bf 31} (1985) 1733.

\bibitem{composite2}
K.~Choi and J.~E. Kim {\em Phys. Rev. D} {\bf 32} (1985) 1828.

\bibitem{composite3}
M.~B. Gavela, M.~Ibe, P.~Quilez, and T.~T. Yanagida {\em Eur. Phys. J. C} {\bf
  79} (2019), no.~6 542, [\href{http://arxiv.org/abs/1812.08174}{{\tt
  arXiv:1812.08174}}].

\bibitem{srednicki}
M.~Srednicki {\em Nucl. Phys. B} {\bf 260} (1985) 689--700.

\bibitem{nardi}
L.~Di~Luzio, M.~Giannotti, F.~Mescia, E.~Nardi, S.~Okawa, and G.~Piazza {\em
  Phys. Rev. D} {\bf 108} (2023), no.~11 115004,
  [\href{http://arxiv.org/abs/2305.11958}{{\tt arXiv:2305.11958}}].

\bibitem{wgc}
N.~Arkani-Hamed, L.~Motl, A.~Nicolis, and C.~Vafa {\em JHEP} {\bf 06} (2007)
  060, [\href{http://arxiv.org/abs/hep-th/0601001}{{\tt hep-th/0601001}}].

\bibitem{Harlow:2022gzl}
D.~Harlow, B.~Heidenreich, M.~Reece, and T.~Rudelius
  \href{http://arxiv.org/abs/2201.08380}{{\tt arXiv:2201.08380}}.

\bibitem{ibanez}
L.~E. Ibanez and A.~M. Uranga, {\em {String theory and particle physics: An
  introduction to string phenomenology}}.
\newblock Cambridge University Press, 2, 2012.

\bibitem{nilles}
H.~P. Nilles {\em Phys. Rept.} {\bf 110} (1984) 1--162.

\bibitem{string_instanton1}
M.~Dine, N.~Seiberg, X.~G. Wen, and E.~Witten {\em Nucl. Phys. B} {\bf 278}
  (1986) 769--789.

\bibitem{string_instanton2}
R.~Blumenhagen, M.~Cvetic, S.~Kachru, and T.~Weigand {\em Ann. Rev. Nucl. Part.
  Sci.} {\bf 59} (2009) 269--296, [\href{http://arxiv.org/abs/0902.3251}{{\tt
  arXiv:0902.3251}}].

\bibitem{Balasubramanian:2005zx}
V.~Balasubramanian, P.~Berglund, J.~P. Conlon, and F.~Quevedo {\em JHEP} {\bf
  03} (2005) 007, [\href{http://arxiv.org/abs/hep-th/0502058}{{\tt
  hep-th/0502058}}].

\bibitem{rs}
L.~Randall and R.~Sundrum {\em Phys. Rev. Lett.} {\bf 83} (1999) 3370--3373,
  [\href{http://arxiv.org/abs/hep-ph/9905221}{{\tt hep-ph/9905221}}].

\bibitem{Cicoli:2012sz}
M.~Cicoli, M.~Goodsell, and A.~Ringwald {\em JHEP} {\bf 10} (2012) 146,
  [\href{http://arxiv.org/abs/1206.0819}{{\tt arXiv:1206.0819}}].

\bibitem{Cicoli:2021gss}
M.~Cicoli, V.~Guidetti, N.~Righi, and A.~Westphal {\em JHEP} {\bf 05} (2022)
  107, [\href{http://arxiv.org/abs/2110.02964}{{\tt arXiv:2110.02964}}].

\bibitem{Ooguri:2006in}
H.~Ooguri and C.~Vafa {\em Nucl. Phys. B} {\bf 766} (2007) 21--33,
  [\href{http://arxiv.org/abs/hep-th/0605264}{{\tt hep-th/0605264}}].

\bibitem{Reece:2018zvv}
M.~Reece {\em JHEP} {\bf 07} (2019) 181,
  [\href{http://arxiv.org/abs/1808.09966}{{\tt arXiv:1808.09966}}].

\bibitem{lukas}
E.~I. Buchbinder, A.~Constantin, and A.~Lukas {\em Phys. Rev. D} {\bf 91}
  (2015), no.~4 046010, [\href{http://arxiv.org/abs/1412.8696}{{\tt
  arXiv:1412.8696}}].

\bibitem{choi_jeong}
K.~Choi, K.~S. Jeong, and M.-S. Seo {\em JHEP} {\bf 07} (2014) 092,
  [\href{http://arxiv.org/abs/1404.3880}{{\tt arXiv:1404.3880}}].

\bibitem{Cicoli:2021tzt}
M.~Cicoli, A.~Schachner, and P.~Shukla
  \href{http://arxiv.org/abs/2109.14624}{{\tt arXiv:2109.14624}}.

\bibitem{Grimm:2007xm}
T.~W. Grimm {\em JHEP} {\bf 10} (2007) 004,
  [\href{http://arxiv.org/abs/0705.3253}{{\tt arXiv:0705.3253}}].

\bibitem{McAllister:2008hb}
L.~McAllister, E.~Silverstein, and A.~Westphal {\em Phys. Rev. D} {\bf 82}
  (2010) 046003, [\href{http://arxiv.org/abs/0808.0706}{{\tt
  arXiv:0808.0706}}].

\bibitem{Candelas:1987kf}
P.~Candelas, A.~M. Dale, C.~A. Lutken, and R.~Schimmrigk {\em Nucl. Phys. B}
  {\bf 298} (1988) 493.

\bibitem{Kreuzer:2000xy}
M.~Kreuzer and H.~Skarke {\em Adv. Theor. Math. Phys.} {\bf 4} (2000)
  1209--1230, [\href{http://arxiv.org/abs/hep-th/0002240}{{\tt
  hep-th/0002240}}].

\bibitem{Grimm:2004uq}
T.~W. Grimm and J.~Louis {\em Nucl. Phys. B} {\bf 699} (2004) 387--426,
  [\href{http://arxiv.org/abs/hep-th/0403067}{{\tt hep-th/0403067}}].

\bibitem{Giddings:2001yu}
S.~B. Giddings, S.~Kachru, and J.~Polchinski {\em Phys. Rev. D} {\bf 66} (2002)
  106006, [\href{http://arxiv.org/abs/hep-th/0105097}{{\tt hep-th/0105097}}].

\bibitem{Becker:2002nn}
K.~Becker, M.~Becker, M.~Haack, and J.~Louis {\em JHEP} {\bf 06} (2002) 060,
  [\href{http://arxiv.org/abs/hep-th/0204254}{{\tt hep-th/0204254}}].

\bibitem{kklt}
S.~Kachru, R.~Kallosh, A.~D. Linde, and S.~P. Trivedi {\em Phys. Rev. D} {\bf
  68} (2003) 046005, [\href{http://arxiv.org/abs/hep-th/0301240}{{\tt
  hep-th/0301240}}].

\bibitem{McAllister:2023vgy}
L.~McAllister and F.~Quevedo \href{http://arxiv.org/abs/2310.20559}{{\tt
  arXiv:2310.20559}}.

\bibitem{Conlon:2006tq}
J.~P. Conlon {\em JHEP} {\bf 05} (2006) 078,
  [\href{http://arxiv.org/abs/hep-th/0602233}{{\tt hep-th/0602233}}].

\end{thebibliography}\endgroup

\end{document}